\documentstyle[epsf]{pasjemu}

\newcommand{\vol}{51}
\newcommand{\no}{3}
\newcommand{\lett}{}
\newcommand{\spage}{1}
\newcommand{\rdate}{1999 April 6}
\newcommand{\adate}{1999 April 26}

\newcommand{\Nv}   {{N\,{\sc v}}}
\newcommand{\Civ}  {{C\,{\sc iv}}}
\newcommand{\Siii} {{Si\,{\sc ii}}}
\newcommand{\Siiv} {{Si\,{\sc iv}}}
\newcommand{\Alii} {{Al\,{\sc ii}}}
\newcommand{\Aliii}{{Al\,{\sc iii}}}
\newcommand{\Mgii} {{Mg\,{\sc ii}}}
\newcommand{\Feii} {{Fe\,{\sc ii}}}
\newcommand{\Feiii}{{Fe\,{\sc iii}}}
\newcommand{\Niii} {{Ni\,{\sc ii}}}
\newcommand{\Niiii}{{Ni\,{\sc iii}}}
\newcommand{\Ciii} {{C\,{\sc iii}]}}
\newcommand{\etal} {{{et~al.\,}}}

\begin{document}

\title{On the Continuum Shape of Broad Absorption Line Quasars}

\author{T. M. {\sc Yamamoto}$^{1,2}$ and
Vladas {\sc Vansevi\v{c}ius}$^{2,3}$
\\[6pt]
$^1$ {\it Institute of Astronomy, The University of Tokyo,
2-21-1 Osawa, Mitaka, Tokyo 181-8588}\\
$^2$ {\it National Astronomical Observatory,
2-21-1 Osawa, Mitaka, Tokyo 181-8588}\\
$^3$ {\it Institute of Physics, Go\v{s}tauto 12,
Vilnius 2600, Lithuania}\\
{\it E-mail (TY): T.Yamamoto@nao.ac.jp}}

\abst{
The continuum shape of 26 broad absorption line (BAL) QSOs is shown to
be virtually identical to that of non-BAL QSOs when the spectra of the
former are dereddened.
A procedure for dereddening of BAL QSO spectra is introduced, based on
measurement of the flux in three bands and determination of the colour
excess in comparison with an unreddened spectrum.
This method allows us to derive the type and amount of extinction.
We show that reddening in high-ionization BAL QSOs is comparatively
small and the range of extinction laws varies from Milky Way, LMC to SMC
type.
On the other hand, low-ionization BAL QSOs seem to experience reddening
which can be attributed to SMC type extinction alone.
Comparison of the colour indices of composite spectra from Francis
\etal (1991) and Zheng \etal (1997) with those derived from individual
BAL QSO spectra shows that the Francis \etal (1991) composite might be
a more suitable representative of an unreddened spectrum.
In general, we conclude that the spectral energy distribution of normal
QSOs can be used as an incident continuum shape for BAL QSOs.
}

\kword{Interstellar: dust --- Interstellar: extinction ---
Quasars: general --- Quasars: absorption lines ---
Ultraviolet: galaxies}

\maketitle
\thispagestyle{headings}

\section{Introduction}

In optically selected QSO samples, $\sim$10\% are broad absorption
line (BAL) QSOs (Weymann \etal 1991).
BAL QSOs can be subdivided into high- and low-ionization (hi-BALs and
lo-BALs) where hi-BALs have absorptions from high ionization species
(\Nv, \Civ, and \Siiv) and lo-BALs, which make up only $\sim$10\% of
the BAL QSO population, have additional absorptions from low-ionization
species (\Aliii, \Alii, and sometimes \Mgii).
In some cases lo-BALs also have \Feii\ and \Feiii\ absorptions (iron
lo-BALs).

Becker \etal (1997) reported the discovery of two new iron lo-BALs from
radio-selected samples (FIRST survey; Becker \etal 1995)
with one of the objects (1556+3517) showing severe extinction in the
spectrum.
Goodrich (1997) pointed out that a significant number of BAL QSOs are
missed in optically selected samples due to attenuation of the continuum
and that the true fraction of BAL QSOs might be as high as 30\% instead
of 10\%.
This could mean that these heavily reddened objects are more common
among QSOs than previously thought.

The continuum shape of BAL QSOs is often flat or distorted compared to
that of non-BAL QSOs (Weymann \etal 1991).
Sprayberry and Foltz (1992) showed that the composite spectra of high-
and low-ionization BAL QSOs can be brought into coincidence by
dereddening the lo-BAL composite spectrum {\em relative} to the hi-BAL
composite with appropriate extinction laws.
This method shows only the tendency for lo-BALs to be redder than hi-BALs
on average.

The present work describes a systematic method for estimating the amount
of extinction for individual hi- and lo-BAL spectra.
The systematic method for deriving extinction for all BAL QSO types
allows a homogeneous comparison of the whole data set, which can also be
extended to non-BAL QSO spectra.
Section~2 describes the observational data and section~3 will introduce
the dereddening procedure for BAL QSOs followed by results and
discussions in section~4.
Conclusions are shown in section~5.

\begin{table*}[ht]
\begin{center}
Table~1.\hspace{4pt}Main parameters of the program quasars.
\end{center}
\vspace{6pt}
\tabcolsep 3pt
\begin{tabular*}{\textwidth}{@{\hspace{2\tabcolsep}
\extracolsep{\fill}}p{3cm}ccccccc@{\qquad\qquad}}
\hline
\hline\\[-2mm]
Object &
$z_{\rm em}$ & \multicolumn{1}{c}{BAL type} &
\multicolumn{1}{c}{$F_{1750}/F_{2100}$} &
\multicolumn{1}{c}{$F_{2100}/F_{3000}$} &
Extinction type & Ref.\\[2mm]
\hline\\[-3mm]
0004$+$0147     \dotfill & 1.706 & hi--lo & 1.16 & 1.68 & silicate/SMC & b \\[1mm]
0018$+$0047     \dotfill & 1.816 & hi     & 1.37 & 1.72 & --           & b \\[1mm]
0025$-$0151     \dotfill & 2.072 & hi     & 1.25 & 1.51 & SMC-LMC      & a \\[1mm]
0051$-$0019     \dotfill & 1.705 & hi     & 1.23 & 1.31 & SMC-LMC      & b \\[1mm]
0054$+$0200     \dotfill & 1.868 & hi     & 1.28 & 1.39 & LMC          & b \\[1mm]
0059$-$2735     \dotfill & 1.584 & iron   & 1.09 & 1.33 & SMC          & a,e \\[1mm]
0145$+$0416     \dotfill & 2.029 & hi     & 1.22 & 1.44 & SMC-LMC      & a \\[1mm]
0840$+$3633$^*$ \dotfill & 1.225 & iron   & 1.01 & 1.16 & SMC          & c,d,f\\[1mm]
0932$+$5010     \dotfill & 1.914 & hi--lo & 1.30 & 1.75 & --           & a \\[1mm]
1009$+$0222     \dotfill & 1.343 & hi--lo & 1.11 & 1.57 & silicate/SMC & b \\[1mm]
1205$+$1436     \dotfill & 1.629 & hi     & 1.27 & 1.46 & SMC-LMC      & a \\[1mm]
1208$+$1535     \dotfill & 1.956 & hi     & 1.37 & 1.44 & LMC-MW       & a \\[1mm]
1212$+$1445     \dotfill & 1.621 & hi     & 1.22 & 1.44 & SMC-LMC      & a \\[1mm]
1216$+$1103     \dotfill & 1.615 & hi     & 1.41 & 1.42 & LMC-MW       & a \\[1mm]
1228$+$1216$^*$ \dotfill & 1.408 & hi     & 1.32 & 1.44 & LMC          & b \\[1mm]
1239$+$0955     \dotfill & 2.014 & hi     & 1.27 & 1.96 & --           & a \\[1mm]
1303$+$3080$^*$ \dotfill & 1.760 & hi     & 1.42 & 1.47 & LMC-MW       & a \\[1mm]
1331$-$0108$^*$ \dotfill & 1.867 & lo     & 0.88 & 1.38 & silicate     & a \\[1mm]
1556$+$3517$^*$ \dotfill & 1.487 & iron   & 0.60 & 0.27 & SMC          & c \\[1mm]
1641$+$4115     \dotfill & 2.005 & hi     & 1.25 & 1.45 & SMC-LMC      & a \\[1mm]
2154$-$2005     \dotfill & 2.028 & hi     & 1.38 & 1.68 & --           & a \\[1mm]
2201$-$1834     \dotfill & 1.817 & lo     & 0.94 & 1.10 & silicate     & a \\[1mm]
2202$-$2007     \dotfill & 2.188 & lo     & 1.21 & 1.83 & --           & b \\[1mm]
2225$-$0534$^*$ \dotfill & 1.981 & hi--lo & 1.15 & 1.09 & SMC-LMC      & a \\[1mm]
2350$-$0045     \dotfill & 1.626 & hi--lo & 1.28 & 1.53 & SMC-LMC      & a \\[1mm]
2358$+$0216     \dotfill & 1.854 & lo     & 0.78 & 1.08 & silicate     & b \\[1mm]
\hline
\end{tabular*}\\[1mm]
a:~Weymann \etal\ (1991);
b:~Korista \etal\ (1993);
c:~Brotherton \etal\ (1997);
d:~Becker \etal\ (1997)
e:~Ogle (1997, private communication);
f:~Wills (1998, private communication).\\
$^*$ Spectrum shown in figure~5.
\end{table*}

\section{Observational Data}

The observational data consist of a total of 26~BAL QSO spectra; 15
objects are from Weymann \etal (1991), 8 from Korista \etal (1993), 2
from Brotherton \etal \linebreak (1997), 1 from Ogle (1997, private
communication) and parts of spectra from Becker \etal (1997) and Wills
(1998, private communication) have kindly been made available by the
authors.
From the sample we chose only those spectra which had spectral coverage
of all three bands defined below in order to do the colour
measurements.

We divide the quasars into hi- and lo-BAL types (hi and lo in table 1),
similar to the classification in Weymann \etal (1991).
However, the classification for hi- and lo-BALs is somewhat ambiguous
and for those cases where it is not clear which group an object belongs
to we indicated this with ``hi--lo''.
Additionally we use the classification of the iron lo-BALs (iron in
table 1) which was introduced by Becker \etal (1997).
Table~1 (columns 1, 2, and 3) lists the names of the program BAL
QSOs, the redshifts and the classifications of the objects.

\newpage

\section{Dereddening Procedure}

The extinction laws used here are the Milky Way (MW), Large Magellanic
Cloud (LMC), and Small Magellanic Cloud (SMC) from Pei (1992), the
attenuation curve from Calzetti \etal (1994) and theoretical extinction
curves for astronomical silicate with different grain sizes from Laor and
Draine (1993).

\subsection{The Unreddened Continuum}

Sprayberry and Foltz (1992) established the idea for the existence of
dust extinction in BAL QSOs by showing that lo-BALs are reddened
relative to hi-BALs.
We take a more general approach by comparing BAL QSOs with a
representative non-BAL QSO sample.
By comparing BAL QSO spectra with a single spectral shape we assume that
all BAL QSOs are reddened relative to the representative spectrum, which
can be understood as an ``intrinsic quasar spectrum''.
We refer to subsection~4.1 for discussion of the intrinsic quasar spectrum.

For the present investigation the following spectra are taken into
consideration for determining the continuum shape of an unreddened QSO
spectrum:
(a) the composite spectrum from Francis \etal (1991; hereafter FSED),
(b) the composite spectrum Zheng \etal (1997; hereafter ZSED), and
(c) an artificially steepened FSED (hereafter FSED-plus) with the
spectral index (defined by $F_\nu\sim\nu^\alpha$) corresponding to
$\alpha=-1$ between 1250 and 2100~\AA\ and $\alpha=+0.4$ between 2100
and 3100~\AA.
We note that spectra of BAL QSOs are included in the composite spectrum
from Francis \etal (1991) but the final spectral slope of the composite
was chosen to have the median slope and thus the BAL continuum will
hardly affect the final shape.

The spectral slopes of FSED-plus result from the following consideration:
A comparison (e.g. figure~9 in Zheng \etal 1997) shows that FSED and
ZSED have an almost identical continuum shape between 1250~\AA\
and 2100~\AA\ with $\alpha\approx-1$.
However, longward of 2100~\AA\ the Francis \etal (1991) composite
spectrum is harder with $\alpha\approx-0.3$,
whereas the Zheng \etal (1997) remains flat ($\alpha\approx-1$).
Examination of the distribution of spectral slope (Francis \etal
1991, figure~1) shows that $\alpha$ has a wide spread around the
median value of $\alpha=-0.32$.
Thus we can consider $\alpha=-1$ (soft spectrum) as a lower bound and
$\alpha=+0.4$ (hard spectrum) as the upper bound for the spectral index
for wavelengths $> 2100$~\AA.
Our choice of FSED-plus together with the composite spectra FSED and ZSED
should cover most of the range in the spectral index.

For the sake of clarity later in figure 5 we approximate the continuum
shape of the composite spectrum FSED as a broken power-law with
$\alpha=-1$ for 1200~\AA\ $<\lambda<$ 2100~\AA\ and $\alpha=-0.32$ for
$\lambda>$~2100~\AA.
Note that all measurements are carried out on the composite spectra and
not on the approximate power-law.

\subsection{Definition of the Bands}

\begin{figure}[t]
\figurenum{1}
\epsscale{1}
\plotone{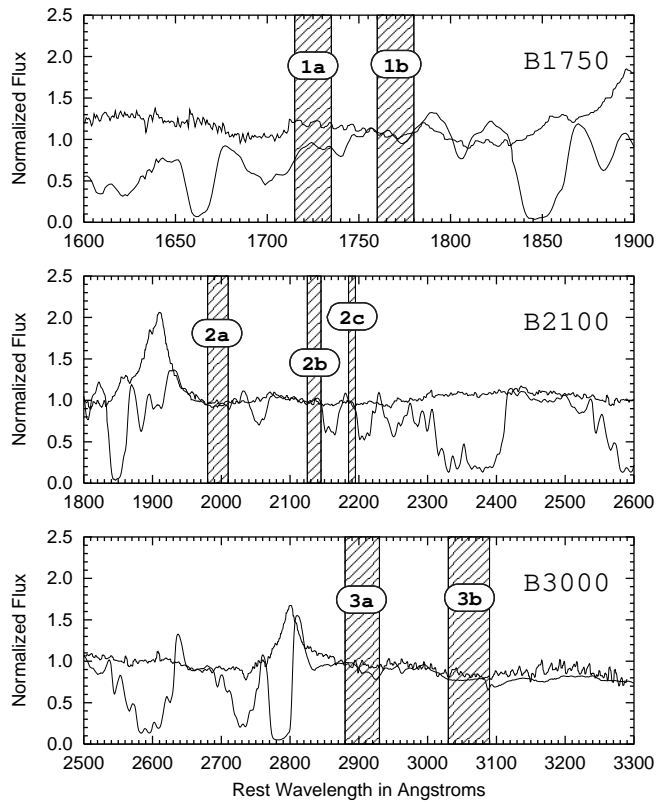}
\caption{
The windows in the bands B1750, B2100, and B3000 are shown.
The two spectra shown are 0840+3633 and 2225$-$0534.
}
\end{figure}

We define three bands in the spectrum at restframe of the quasar around
1750 \AA, 2100 \AA, and 3000 \AA.
Measurement of the continuum level of BAL QSO spectra is difficult,
since BAL spectra have numerous absorption lines of high- and
low-ionization elements that arise in the broad and narrow line region
of the quasar.
Therefore, windows, which seem relatively free from either emission or
absorption, were defined within the bands as follows:
\begin{description}
\label{banddef.page}
\item[B1750:] at 1715--1735~\AA\ and 1760--1780~\AA
\item[B2100:] at 1980--2010~\AA, 2125--2145~\AA, and 2185--2195~\AA
\item[B3000:] at 2880--2930~\AA\ and 3030--3090~\AA
\end{description}
The total coverage of the bands is $\Delta\lambda\approx1250$~\AA\ with
separation of 350~\AA\ and 900~\AA\ between the bands.
This is sufficient to trace extinction effects in the UV region where
the extinction curves rise steeply.
However, the necessity for a wide spectral coverage restricts the number
of objects which can be analyzed.
Note that B2100 is located near to the 2175~\AA\ feature of the Galactic
extinction curve.

Figure 1 shows the position of the windows in the individual bands.
As an example we show two BAL spectra, one which has relatively strong
absorption lines of low and high ionization species (lo-BAL 0840+3633)
and another, which has only high ionization lines (hi-BAL 2225$-$0534).

\begin{figure*}[ht]
\figurenum{2}
\epsscale{1.9}
\plotone{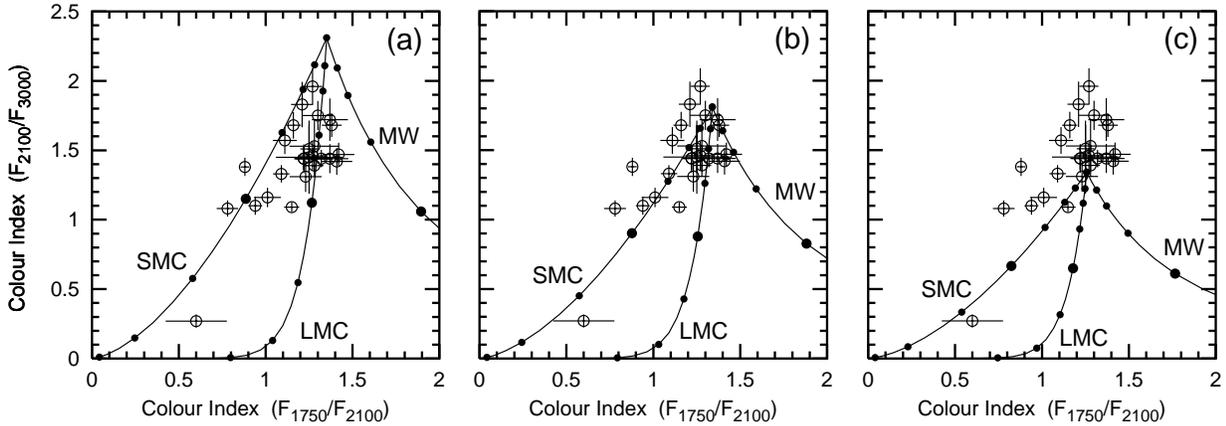}
\caption{
The effect of extinction on the colour indices is shown.
(a) for the case of the artificially steepened Francis \etal (1991)
spectrum FSED-plus defined in subsection~3.1 (b) for Francis \etal
(1991) composite spectrum and (c) for Zheng \etal (1997) composite
spectrum.
The different lines show the effect of extinction when the empirical
extinction laws of the MW, LMC, and SMC are applied.
The type of extinction law is indicated in the figure.
The extinction increases along the lines starting from the filled circle
at the top of the lines which represents the measurement of the
unreddened composite spectrum.
The observations (open circles) are shown with errorbars.
The dots on the lines mark the amount of extinction increasing by a
factor of 2 from dot to dot.
The thick dot on each line indicates $E(B-V)=0.25$.
}
\end{figure*}

The choice of the windows can be summarized as follows:\\
\noindent
{\bf Band around 1750~\AA:} (figure 1, upper panel) This band lies
between \Civ\ and the semiforbidden emission line of \Ciii.
Two windows (1a) at 1715--1735~\AA\ and (1b) 1760--1780~\AA\ cover a
total of 40~\AA.
The strong absorption lines are the singlet \Alii\,$\lambda\,1671$, the
doublets \Siii\,$\lambda\lambda\,1808, 1817,$ and
\Aliii\,$\lambda\lambda\,1855, 1863$.
The \Siii\ line is only $\Delta\lambda=28$~\AA\ away (which corresponds
to a velocity of $v=c\,\Delta\lambda/\lambda\sim4\,000$~km\,s$^{-1}$)
from window (1b) but the blueshift of broad absorption lines of \Siii\
seldom exceed velocities of 4\,000~km\,s$^{-1}$ and thus are unlikely to
affect the measurement in this window.
\Aliii\ shows much stronger BALs.
However it is unlikely for the width of the absorption line to exceed
13\,000~km\,s$^{-1}$ which would diminish measurements in (1b).
The weaker features are absorptions from multiplet lines of \Feii\ and
possibly \Niiii.
In strongly absorbed quasar spectra these weak emissions and absorptions
may have an effect on window (1a).
The emission-like feature around 1790~\AA\ is caused by the \Feii\ uv
191 transition.

\noindent
{\bf Band around 2100~\AA:} (figure 1, middle panel) The second band
lies between \Ciii\ and \Mgii.
Three windows at 1980--2010~\AA, 2125--2145~\AA, and 2185--2195~\AA\
cover a total of 60~\AA.
This definition is close to the one used by Weymann \etal (1991).
Window (2a) is positioned before the rise of the semiforbidden line
emission of \Ciii\,$\lambda\,1909$ on the red side.
The soft rise between (2a) and (2b) with its peak around 2070~\AA\ is
likely caused by emission from the \Feii\ multiplet.
This iron multiplet also shows BALs in some quasars but its absorption
is unlikely to reach window (2a) which is 9\,000~km\,s$^{-1}$ away.
The well-known strong multiplets of \Feii\ uv 1, 64 (around 2600~\AA)
and \Feii\ uv 2, 3, 35, 36 (around 2400~\AA) produce deep blueshifted
BALs in 0840+3633.
The series of three strong absorptions is likely caused by \Feii\
(around 2250~\AA) and possibly \Niii\ (around 2210~\AA\ and 2170~\AA).
Windows (2b) and (2c) are set to avoid these spectral features.

\noindent
{\bf Band around 3000~\AA:} (figure 1, lower panel) The third band is
on the red side of \Mgii.
Two windows at 2880--2930~\AA\ and 3030--3090~\AA\ cover a total of
110~\AA.
The main feature here is the emission lines of
\Mgii\,$\lambda\lambda\,2796, 2803$.
The continuum level around \Mgii\ is raised by numerous \Feii\ multiplets
and the Balmer continuum of hydrogen.
Although the continuum to the red of \Mgii\ seems smooth, there appears
a weak emission feature around 3000~\AA, possibly caused by \Feii, which
varies in strength from object to object.
Window (3a) is placed between \Mgii\ and this emission feature and (3b)
is placed to the red side of the emission.

\subsection{Calibration of the Diagnostic Diagram}

The measurement of the fluxes $F_{1750}$, $F_{2100}$, and $F_{3000}$ in
the bands is obtained by integrating the spectrum over the windows and
dividing it by the total width of the bands, i.e. average flux per \AA.
Although the windows have been carefully chosen, it is important to check
e.g. for overlapping BALs when measuring the bands and to correct the
measure if necessary.

Table~1 (columns 4 and 5) gives the derived colour indices which are
defined as $F_{1750}/F_{2100}$ and $F_{2100}/F_{3000}$.
The two-colour diagnostic diagram (TCDD) is obtained by plotting colour
indices as ($x,y$)-pairs.
The errors (shown in figure~2) are estimated by dividing the maximum
difference of flux in a band with the square root of the number of bins
within the integration interval.

\begin{figure*}[t]
\figurenum{3}
\epsscale{1.9}
\plotone{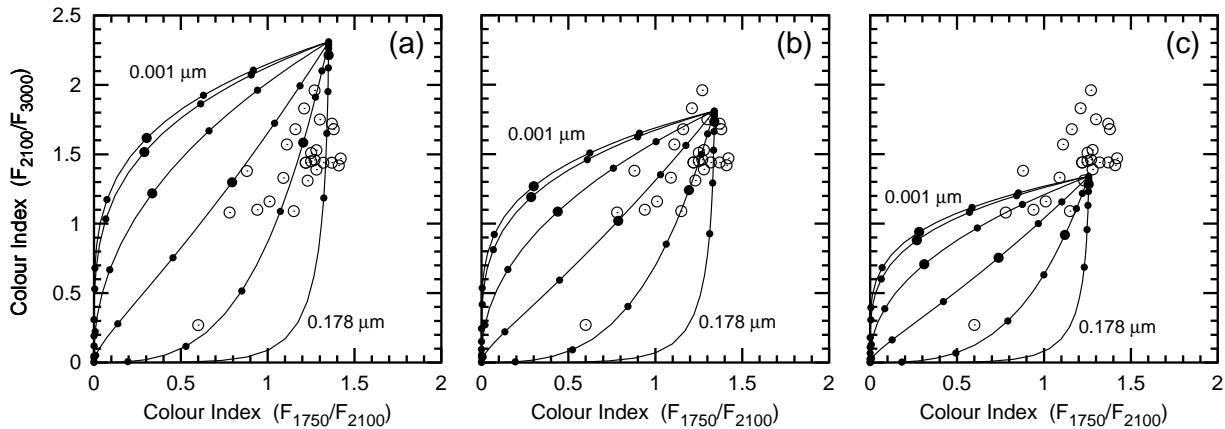}
\caption{
Same as figure~2 but for astronomical silicate with different grain
sizes.
The thick dots indicate $E(B-V)=0.1$.
Only the lines corresponding to the smallest and the biggest grain sizes
are marked.
For clarity, errorbars have been omitted in this figure.
}
\end{figure*}

The effects of reddening with the various extinction laws, including
SMC, LMC, and MW type from Pei (1992) and astronomical silicate from Laor
and Drain (1993), are investigated by applying them to the composite
spectra from Francis \etal (1991) and Zheng \etal (1997).
After reddening the composite spectra the intensities in the bands are
determined with the same procedure as for the observed spectrum.

For simplicity, we use single grain size for the extinction curve of
astronomical silicate where the following sizes are used (given is the
logarithm of grain sizes in $\mu$m): $-$3, $-$2, $-$1.5, $-$1.35,
$-$1.25, and $-$.75.

Numerous multiplet lines of iron-group elements can either depress the
continuum or produce emission features in the bands and influence the
measurements of the colour indices.
Unfortunately, at this stage model calculations of the low-ionization
region cannot provide hints of whether the continuum level in the bands
is lowered or raised.
At present, our definition of the colour indices provides a homogeneous
measurement for all quasars, including lo-BALs, and yields a robust
estimate of the continuum shape.

\section{Results and Discussions}

The extinction of the quasars can be obtained from the TCDD once the
comparison spectrum for the diagram is chosen (see subsection~4.1).
In order to distinguish the type of extinction more accurately we
introduce a classification scheme for the determination of the
extinction type in subsection~4.2.
The information obtained from the TCDD is used to deredden the spectra
and we show examples for six of the program quasars at the end of this
section.

\subsection{The Two-Colour Diagnostic Diagrams}

Figure~2 shows the positions of the program quasars in the TCDD together
with lines produced from FSED-plus (panel~a), FSED (panel~b), and ZSED
(panel~c) applying empirical extinction laws.

Figure~3 shows the same program quasars as in figure~2 but this time
lines are produced by applying theoretical extinction curves of
astronomical silicate with different grain sizes to FSED-plus, FSED, and
ZSED (panels~a, b, and c respectively).
By comparing figures~2 and 3 it becomes apparent that intermediate and
large grain sizes mimic the lines in figure~2 that are produced by SMC
and LMC extinction law respectively.
The region above the line produced by SMC in panel~b of figure~2 may be
explained by reddening FSED using extinction curves of astronomical
silicate with small grain sizes.

The first thing to notice in figure~2 is that the bulk of observations
have $F_{2100}/F_{3000}$ values which are higher than those derived from
ZSED at the top where the curves join (figure~2 panel~c).
We attribute this to a poor representation of the flux around 3000~\AA\
where the composite is derived from only a few quasars.
The position of FSED in the TCDD, on the other hand, is at one of the
bluest positions in the diagram (panel~b).
FSED-plus is further to the blue in the diagram (panel~a) as expected
since we chose a value resulting in the hardest (bluest) spectrum in our
comparison.

\begin{figure*}[t]
\figurenum{4}
\epsscale{1.35}
\plotone{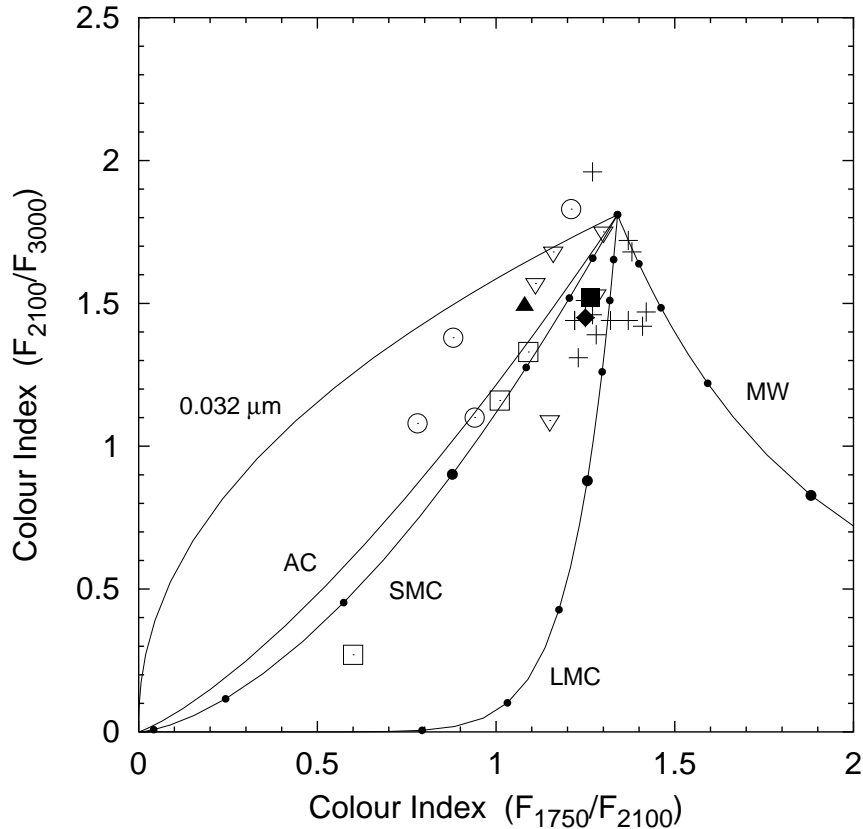}
\caption{
The effect of extinction on the Francis \etal (1991) composite spectrum
for the empirical extinction curves MW, LMC, SMC, and attenuation curve
(AC) (Calzetti \etal 1994) is shown.
For the discussion is shown the line produced by astronomical silicate
with grain size ${\rm a}=0.032 \mu$m.
The different symbols indicate hi-BALs (plus signs), lo-BALs (open
circles), iron lo-BALs (open squares), and the quasars labeled as
``hi--lo'' in table~1 (open triangles).
For comparison we have included the position of the average spectra from
Weymann \etal (1991) non-BAL (filled diamond), lo-BAL (filled triangle),
and hi-BAL (filled square).
The filled circles on the lines mark the amount of extinction as
described in figure~2.
}
\end{figure*}

Nonetheless, the upper sequence of observations (see e.g. figure~2
panel~b, to the left of the SMC-line) implies the existence of a yet
bluer quasar spectrum.
Such an {\em intrinsic spectrum} would have a spectral slope which is
harder than the Francis \etal (1991) composite.
The artificially hardened FSED-plus is only one way to represent the
intrinsic spectrum, however, it should be reminded that FSED-plus is
only a hypothetical spectrum whereas FSED and ZSED are derived from
observations and is therefore less desirable.
Another method of obtaining the intrinsic spectrum of quasars is to take
the steepest spectra from a quasar survey and construct a composite
spectrum.
If such an intrinsic spectrum exists it would mean that almost all
quasar spectra observed are reddened by dust.

\subsection{Determination of the Extinction Types}

From the discussion above we conclude that the ZSED and FSED-plus are less
suitable for representing the intrinsic shape of BAL QSOs.
Therefore, the determination of the amount and type of extinction will be
carried out using FSED.

Figure~4 shows the TCDD with lines for the case of Francis \etal (1991)
composite spectrum when empirical extinction laws are applied.
From the position of the curves that are produced using the SMC, LMC, and
MW extinction laws we define the following classification for the
extinction type:
SMC, SMC-LMC, LMC, and LMC-MW.
Additionally we define the ``silicate'' type for objects on the left of
the SMC-line.
We subdivide the regions between SMC, LMC, and MW and decide upon the
positions of the objects in the TCDD between the extinction types.
In cases where it does not seem clear whether the spectrum has silicate
or SMC type extinction this was marked with ``silicate/SMC'' in table~1.
Note that $F_{2100}/F_{3000}$ mainly provides information about the
amount of dust while $F_{1750}/F_{2100}$ generally distinguishes between
extinction types.

The lo-BALs with iron absorptions (0059$-$2735 and 0840$+$3633; open
squares in figure 4) lie on the SMC-line.
Within its errors, even the iron lo-BAL 1556$+$3517, which suffers
severe extinction, lies near the SMC-line (see panel~b of figure~2).
An alternative solution for iron lo-BALs gives astronomical silicate
with grain size of about 0.05~$\mu$m.
However, better agreement of the continuum level at shorter wavelengths
between \Civ\,$\lambda\,1549$ and \Siiv\,$\lambda\,1394$ is achieved
when the spectra are dereddened with the SMC extinction law.

The lo-BALs without iron absorption (1331$-$0108, 2202$-$2007 and
2358+0216; open circles in figure 4) have a peculiar continuum shape
which neither of the empirical extinction laws can explain.
In this case astronomical silicate with small grain size of about 0.03
to 0.04~$\mu$m best reproduces the extinction in all three bands.

The hi--lo and hi-BALs, on the other hand, have a wider distribution near
to the ``origin'' of the curves in the TCDD.
However their small amount of extinction makes it difficult to uniquely
distinguish the type of extinction.

Table~1 (column 6) shows the determined extinction type for each object.
For those objects which have positions above the silicate-line (grain
size 0.001 $\mu$m) no extinction type could be derived and they have
been marked with ``--'' in table~1.

Remarkable is that most of the observations in the TCDD do not lie on the
curve produced with the MW extinction law.
There are only two quasars that have extinction type LMC-MW in our
sample.
Since $F_{2100}$ is very sensitive to the 2175~\AA\ feature, we can
conclude from the lack of MW type extinction objects that the 2175~\AA\
bump is virtually absent in BAL QSOs.
For the same reason, theoretical extinction curves of graphite grains
(Laor, Drain 1993), which have pronounced features at around 2200~\AA,
are unsuitable for dereddening of BAL QSO spectra.

There is a clear division of the determined extinction laws (table~1)
between lo-BALs and hi-BALs.
This can be due to different viewing angles amongst various types of BAL
QSOs (Yamamoto 1998) which means that the areas where the extinction
occurs are in separate regions where different dust composition, grain
size distribution, and physical conditions might affect the extinction
law.
It might also be due to a strong selection effect of the lo-BALs since
only a few objects are known.

\subsection{Extinction with Spatial Distribution of Dust}

The spatial distribution of dust is regarded as important for
determining the effect of extinction on the spectrum in the case of
galaxies (e.g. Witt \etal 1992; Calzetti \etal 1994).
In the case of quasars, however, there is only one continuum source with
most of the continuum-forming region being too hot for dust to survive.
Therefore, the uniform dust-screening model commonly used for stars
might be the most suitable configuration of dust for investigations of
extinction effects in quasars.

To demonstrate the robustness of the TCDD in the case of more arbitrary
spatial distribution of the dust in extended regions, we apply the
attenuation curve (AC) from Calzetti \etal (1994) to FSED.
Figure~4 illustrates that the AC-line lies on the left side close to the
SMC-line and shows that a more complicated configuration than simple
screening does not significantly affect the shape of the lines in the
diagram.
We expect that more realistic treatment of the dust distribution around
quasars and solution of the radiative transfer problem are unlikely to
endanger our assumption of the dust screen model where the absorption
occurs outside the BAL region.

\subsection{Dereddening of Quasar Spectra}

The method employed by Sprayberry and Foltz (1992) to show evidence for
extinction in BAL QSOs used only composite spectra from Weymann \etal
(1991).
We show in our diagram (figure~4) that the composite spectra they used,
namely the hi- and lo-BALs, are {\em both} affected by dust.
This means that estimation of the extinction in lo-BALs was done with
respect to the intrinsically reddened composite spectrum of hi-BALs and
can only be regarded as a hint for dust extinction in quasars.

In our case the type and amount of intrinsic reddening of BAL QSOs are
obtained from the TCDD by looking for the observed points' nearest
positions to the lines produced by the extinction laws.
Since we apply only standard extinction laws (SMC, LMC, MW, and
astronomical silicate) we assign, for simplicity, one type for the cases
where positions in the TCDD indicate an intermediate type of extinction.

Several trials of the dereddening process were made for each object
with the suggestion offered by the TCDD as first guess.
In most cases, the extinction predicted by the TCDD gave satisfactory
results.
Then the spectra in the region around \Civ\ and \Siiv\ were examined for
the quality of the fit and the amount of extinction slightly readjusted
by eye to match the continuum level between \Civ\ and \Siiv.
A more systematic fit would be possible if another band could be defined
in this region.
However, the continuum in this region is subjected to the \Civ\ broad
absorption line, whose maximal width strongly varies from object to
object.
This makes it impossible to define a band between \Civ\ and \Siiv\ that
would serve for all BAL QSOs as a measure of the continuum level.

\subsection{Examples of Dereddened Spectra}

Six objects are chosen (marked with asterisk in table~1) for
presentation of the dereddened spectra as follows:
the object with the highest extinction (1556$+$3517) to show that the
recovery of the spectrum works correctly for extreme cases;
at intermediate extinction, representatives of various extinction types
are chosen (2225$-$0534, 0840$+$3633, and 1331$-$0108, which we have
determined to be LMC, SMC, and silicate type respectively) to show that
spectra dereddened with the predicted extinction types indeed recover
the overall shape of the spectrum;
two are taken from the region with the highest population density in
TCDD (1228$+$1216 and 1303$+$3080) with one of the objects near to the
MW-line in figure~4.

Figure~5 shows dereddened and observed spectra together with the
corresponding power-law continua.
Numerous emission lines significantly change the local shape of the
quasar continuum which cannot be presented with the power-law and
therefore we show FSED in panel (a) as a guide of the continuum level.

\begin{figure*}[t]
\figurenum{5}
\epsscale{1.32}
\plotone{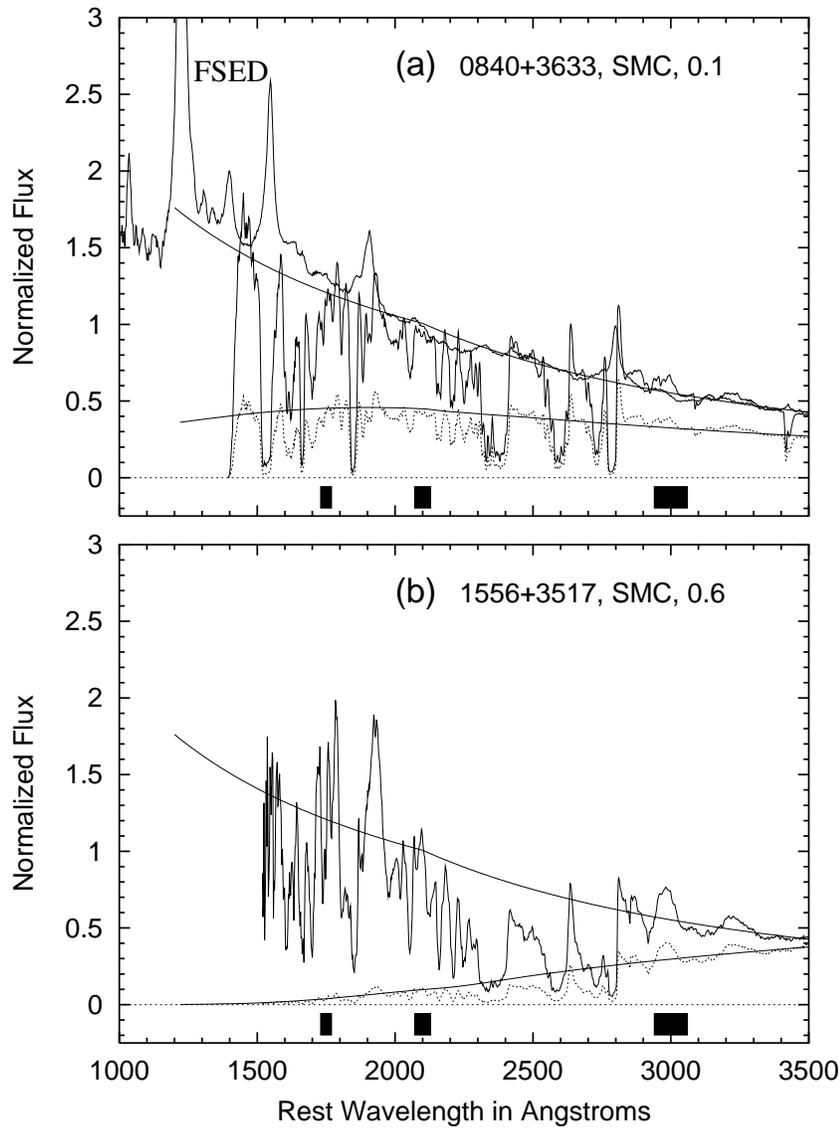}
\caption{
Comparison of dereddened spectra (solid spectrum) with observed
spectra (dotted line).
The continua shown are the power-law representation of the Francis
\etal (1991) composite spectrum without (upper smooth line) and with
(lower smooth line) reddening.
The spectra are normalized at 2100~\AA\ for the unreddened power-law.
The object name, type of extinction law, and $E(B-V)$ are indicated.
The positions of the bands where the flux was measured are indicated as
black rectangles at the bottom of each panel.
Panels (a) and (b) show expamles for SMC type extinctions.
In panel (a) of this figure we have over-plotted the composite spectrum
from Francis \etal (1991).
In panel (b) the observed spectrum and the reddened continuum have been
scaled by a factor of 15 since they would be almost flat in the present
scale.
}
\end{figure*}

\begin{figure*}[t]
\figurenum{5 (continued)}
\epsscale{1.32}
\plotone{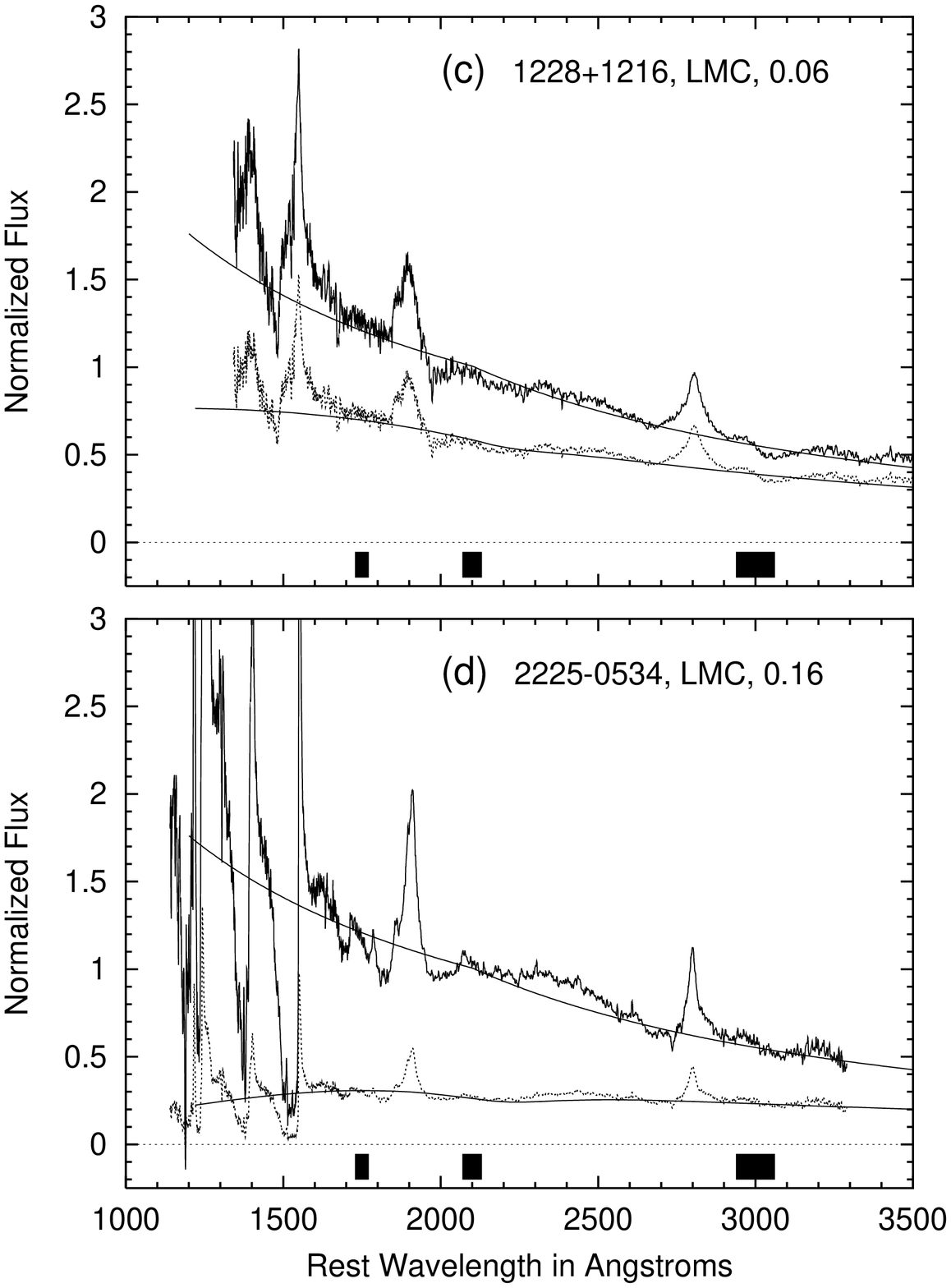}
\caption{
Panels (c) and (d) show expamles for LMC type extinctions.
}
\end{figure*}

\begin{figure*}[t]
\figurenum{5 (continued)}
\epsscale{1.32}
\plotone{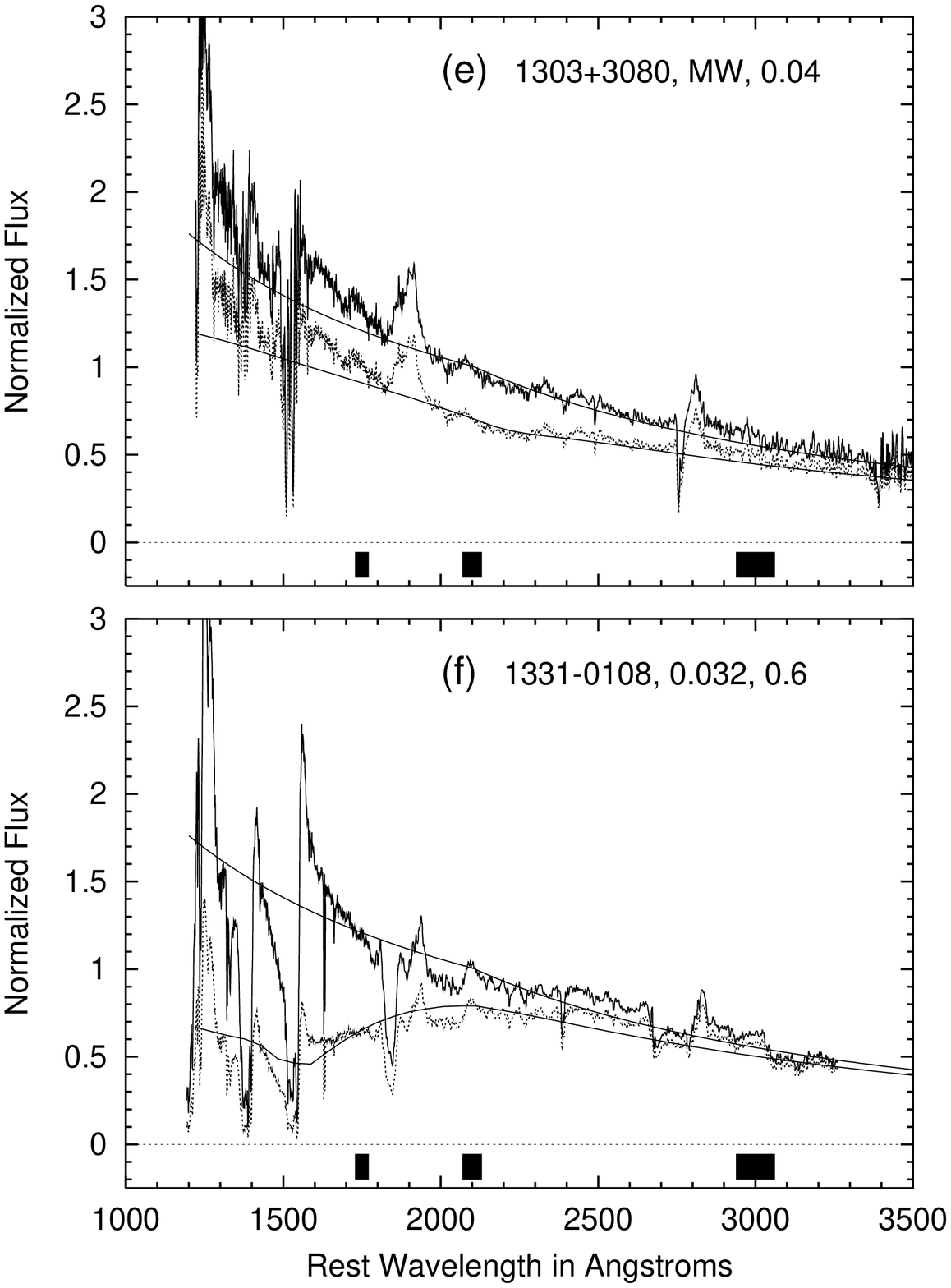}
\caption{
Panel (e) shows an example for MW type extinction.
In the case of astronomical silicate (f) the grain size in $\mu$m is
given instead of the type of extinction law.
}
\end{figure*}

The iron lo-BAL 0840$+$3633 in panel (a) was dereddened with SMC type
extinction.
The successful recovery of the spectrum shows that strong \Mgii\ and
\Feii\ absorption lines do marginally affect the measurements of the
flux.
It is striking to see that even the heavily extinct spectrum of
1556$+$3517 (panel b) can be recovered with the extinction values
determined from the TCDD.
The ratio of the flux in B2100 in this case between observed and
dereddened spectrum is a factor of $\sim60$.
For these two objects adjustment of $E(B-V)$ to values slightly smaller
than suggested by the TCDD was necessary.

The dereddened spectrum of the weakly extinct hi-BAL 1228$+$1216 (panel
c) almost exactly matches FSED.
The significantly stronger reddened hi-BAL 2225$-$0534 (panel d) for
which the LMC type extinction curve was applied also is in good
agreement when compared with FSED.
Note that FSED is only shown in panel (a) of figure~5.

Dereddening of the hi-BAL 1303$+$3080 (panel e), which was classified as
LMC-MW, with a MW type extinction law gave somewhat better results than
did the LMC type.
In the TCDD (figure~4) this object is closest to the MW-line for which
measurable extinction could be derived.
We show one example for astronomical silicate type extinction from our
sample --- 1331$-$0108 (panel f).
In this case, dereddening with the extinction values derived from the
TCDD also leads to satisfactory results.
We note that dereddening of 1331$-$0108 with SMC type extinction law, as
figure~2 (panel~a) would suggest, cannot recover the region around \Civ\
whereas the steep change in the silicate extinction curve in this region
gives a better fit.

In general, better fits were achieved with the empirical extinction laws.
The only exceptions from the whole quasar sample were 1331$-$0108,
2202$-$2007, and 2358+0216, for which only astronomical silicate with a
grain size smaller than for the other objects could recover the
hump-like shape of the continuum at around 1750~\AA.

\section{Conclusions}

26 BAL QSOs have been analyzed in this paper for which we have defined
bands to measure the flux and constructed a two-colour diagnostic
diagram (TCDD).
Based on the TCDD we introduced a classification scheme for the
determination of the extinction type and estimation of the amount of
extinction for individual objects.
Using this information we can deredden observed spectra with standard
extinction laws (SMC, LMC, MW, and astronomical silicate).
From our investigation we draw the following conclusions:

For low-ionization BAL QSOs, the SMC type extinction law is appropriate
for dereddening while the range of extinction laws for high-ionization
BAL QSOs varies from Milky Way, LMC to SMC.

The TCDD also indicates that the Milky Way type extinction curve is not
applicable for most of the BAL QSOs.

The Francis \etal (1991) composite spectrum is more suitable than
Zheng \etal (1997) composite spectrum for the representation of the
continuum shape of BAL QSOs.

The similarity of the shapes of BAL QSOs' dereddened continua to those
of non-BAL QSOs suggests that both types of quasar have the same kind of
continuum source.
This means that a representative non-BAL (normal) QSO spectrum such as
the composite spectrum from Francis \etal (1991) might be appropriate
for determining the continuum shape of incident radiation for the broad
line region.

There is a hint of the existence of an ``intrinsic'' continuum shape of
quasars that is bluer than the Francis \etal (1991) composite spectrum.
This would imply that all quasars observed might have spectra that are
reddened by dust.

\par\vspace{1pc}\par

We are grateful to M.~S.~Brotherton, P.~M.~Ogle, B.~J.~Wills, and
R.~J.~Weymann for making available the observational data.
We also like to thank P.~J.~Francis for having made available the
composite spectrum from Francis \etal (1991).
For valuable comments on the manuscript we thank E.~J.~Wampler,
V.~Korchagin, T.~Nakajima, A.~Ku\v{c}inskas, and M.~Iye.
Finally, we would like to thank Y.~McLean and T.~Hoffmann for careful
reading of the manuscript.
This work was supported by the Japan Society for the Promotion of
Science.

\section*{References}
\re Becker R.H., Gregg M.D., Hook I.M., McMahon R.G., White R.L.,
Helfand D.J.~1997, ApJ 479, L93
\re Becker R.H., White R.L., Helfand D.J.~1995, ApJ 450, 559
\re Brotherton M.S., Tran H.D., van Breugel W., Dey A.,
Antonucci R.~1997, ApJ 487, L113
\re Calzetti D., Kinney A.L., Storchi-Bergmann T.~1994, ApJ 429, 582

\vfill
\pagebreak

\re Francis P.J., Hewett P.C., Foltz C.B., Chaffee F.H.,\linebreak
Weymann R.J., Morris S.L.~1991, ApJ 373, 465 (FSED)
\re Goodrich R.W.~1997, ApJ 474, 606
\re Korista K.T., Voit G.M., Morris S.L., Weymann R.J.~1993, ApJS 88,
357
\re Laor A., Draine B.T.~1993, ApJ 402, 441
\re Pei Y.C.~1992, ApJ 395, 130
\re Sprayberry D., Foltz C.B.~1992, ApJ 390, 39
\re Weymann R.J., Morris S.L., Foltz C.B., Hewett P.C.~1991, ApJ
373, 23
\re Witt A.N., Thronson H.A. Jr, Capuano J.M. Jr~1992, ApJ 393, 611
\re Yamamoto T. M.~1998, PhD Thesis, The University of Tokyo
\re Zheng W., Kriss G.A., Telfer R.C., Grimes J.P., Davidsen
A.F. 1997, ApJ 476, 469 (ZSED)
\label{last}

\end{document}